# Strain-magneto-optics in CoFe$_2$O$_4$: magneto-absorption in Voight geometry

Yu.P.Sukhorukov*, A.V.Telegin, N.G.Bebenin, S.V.Naumov, A.P.Nossov

M.N. Miheev Institute of Metal Physics of Ural Branch of RAS, 620108, Yekaterinburg, Russia
*e-mail: suhorukov@imp.uran.ru

The infrared optical, magneto-optical and magnetostrictive properties of CoFe$_2$O$_4$ single crystal are considered. The magneto-transmission and magneto-reflection of natural light in magnetostrictive CoFe$_2$O$_4$ spinel are studied in the Voight experimental geometry. These magneto-optical effects are very high and associate with a change of the fundamental absorption edge and impurity absorption bands under magnetic field. It is presented the effects strongly depend on both the magnitude and orientation of magnetic field relative to the crystallographic axes of the crystal. The clear connection between magneto-absorption of light in the infrared spectral range and magnetostriction of CoFe$_2$O$_4$ spinel is established. The contribution of magnetostriction to the magnetic anisotropy constant of the CoFe$_2$O$_4$ crystal is shown to be abnormally great.

## 1. INTRODUCTION

Straintronics is a field of spintronics, which studies variation of physical properties of materials associated with elastic deformations arising under the influence of magnetic and/or electric fields[1,2,3]. The variety of magneto-optical effects caused by magnetostriction observed both in polarized light[4,5,6] and in natural light[7] allows one to conclude about the formation of a special branch of straintronics that has been called recently as "strain-magneto-optics"[8]. First observation of the magneto-reflection effect in natural light in the CoFe$_2$O$_4$ ferrimagnetic spinel with strong magnetostriction has been reported in Refs.7,8. It has been showed that field dependences of magneto-reflection, defined as $\Delta R/R=(R_H-R_0)/R_0$, where $R_H$ and $R_0$ are the specular reflection coefficients in the presence and absence of a magnetic field, respectively, correlate well with magnetostriction. Magneto-transmission of CoFe$_2$O$_4$, defined as $\Delta t/t=(t_H-t_0)/t_0$, where $t_H$ and $t_0$ are the light transmission coefficients in the presence and absence of a magnetic field, respectively, has been studied only in the Faraday geometry of the experiment, i.e. for magnetic field normal to the crystal plane[9]. The strong correlation between the field dependences of the magneto-transmission and magnetostriction of the crystal has been found. In addition, the contribution of Faraday rotation due to the partial polarization of light by an optical system to the magneto-transmission of light has been estimated.

In magnetically ordered materials, both $\Delta R/R$ and $\Delta t/t$ can have high values in the infrared (IR) spectral region, where linear magneto-optical Kerr and Faraday effects are much weaker (see

Refs.4,5,6,7,8,9 and references therein). The study of magneto-transmission in the Faraday experimental geometry – especially in the case of materials with strong magnetostriction – is complicated by the "parasitic" mechanical stresses arisen upon application of magnetic field. The stresses result in poorly controlled deformation of a sample and, therefore, misrepresentation of the data obtained. In contrast, in the Voigt geometry, when magnetic field is applied in the plane of a sample, such stresses are minimal. In addition, in the Voight geometry it is convenient to obtain information on the dependence of magneto-optical effects on the orientation of a magnetic field relative to the crystallographic axes of a crystal.

In this paper we present the observation of the optical and magneto-optical properties of $CoFe_2O_4$ magnetostrictive single crystals in the Voight experimental geometry. We show the existence of noticeable magneto-transmission and magneto-reflection effects in natural light at room temperature in relatively weak magnetic fields. The direct correlation between magneto-optical and magneto-elastic properties of the $CoFe_2O_4$ crystals is observed. The abnormally large contribution of magnetostriction to the magnetic anisotropy constant is obtained. The observed magneto-optical phenomena are explained by the magnetic-field induced deformation of the electronic structure of spinel. For comparison, we used some data for the $Hg(Cd)Cr_2Se_4$ ferromagnetic spinel with small magnetostriction coefficient.

## 2. SAMPLES AND EXPERIMENTAL TECHNIQUE

The $CoFe_2O_4$ ferrimagnetic spinel is characterized by high transparency in the IR spectral region, a high Curie temperature of $T_C = 812$ K, and strong magnetostriction. $CoFe_2O_4$ single crystals investigated here were grown using crucible-free zone melting with radiation heating[10]. The lattice parameter was found to be $a_0$=8.38 Å (inverted spinel structure, $Fd3m$ space group[11]), which is close to $a_0$=8.39 Å [11]. By using EDAX, it has been confirmed that the samples are single-phase and their chemical composition corresponds to the $CoFe_2O_4$ formula unit. Detailed description of the samples can be found in Ref.7.

Magnetization measurements were performed with the LakeShore 7400 vibration magnetometer in magnetic field $H$ of up to 17 kOe at room temperature. Magnetostriction was measured by the tensometric method on the (001) oriented plate-shaped samples with an area of 10x10 $mm^2$ and thickness of $d = 400$ μm. During all experiments, the magnetic field was directed along the plane of the samples.

All optical measurements were carried out by using the prism monochromator in magnetic fields of up to $H = 7.5$ kOe. The coefficient of specular reflection of light was determined as $R=I_S/I_{Al}$, where $I_S$ and $I_{Al}$, are the intensity of unpolarized light reflected from the sample and from the Al mirror, respectively. The surface roughness of the polished samples was less than 1 μm. Such the roughness is sufficient for measurement of $R$ in the infrared spectral region at wavelengths $\lambda > 1$ μm. The relative error in the determining $R$ was 0.2%. The coefficient of light transmission determined as $t = Y_S / Y_O$, where $Y_S$ and $Y_O$ are the intensities of the transmitted and incident light, was measured for unpolarized light at an angle of incidence of 7° relative to sample normal. It was found that in the wavelength range of $2.5 \leq \lambda \leq 5$ μm the transmittance was $t > 10$ %. Thus, for calculating the absorption coefficient $\alpha(\lambda, H, T)$, it was necessary to take into account double reflection of light, so that

$$\alpha = \left(\frac{1}{d}\right) \ln \frac{(1-R)^2}{t}. \qquad (1)$$

Magneto-absorption was defined as $\Delta \alpha / \alpha = (\alpha_H - \alpha_0)/\alpha_0$, where $\alpha_H$ and $\alpha_0$ are the values of absorption coefficient under application of a magnetic field $H$ and in zero magnetic field, respectively.

## 3. MAGNETIZATION AND MAGNETOSTRICTION

The results of investigations of magnetic field dependences of magnetization $M(H)$ and relative elongation $\Delta l/l$ for the CoFe$_2$O$_4$ single crystal sample are presented in Fig.1. For the magnetic field directed along [100] and [010] axis the coercive force is $H_c = 80$ Oe, saturation magnetization $M$ is equal to 82 emu/g at $H = 17$ kOe, see Fig.1a. These data are close to those presented in Refs.10,12. The $M(H)$ curve measured in $\mathbf{H} \| [110]$ geometry has two steps at $H \sim 0.8$ kOe ($M \sim 40$ emu/g) and at $H = 4$ kOe ($M \sim 78$ emu/g), which are due to slight distortion of the cubic symmetry of the crystal. Magnetic field dependences of $(\Delta l/l)_{100}$ for $\mathbf{H} \| [100]$ and $\mathbf{H} \| [010]$ (see Fig.1b and 1c) are similar to those for typical CoFe$_2$O$_4$ single crystals[13]. However, the value of $(\Delta l/l)_{100}$ exceeds known values for nonstoichiometric and doped crystals[13,14,15].

At room temperature ($T \ll T_C$) the saturation magnetization of CoFe$_2$O$_4$ is practically independent of $H$. In this case, volume magnetostriction is small and can be neglected. Then in a cubic ferromagnet the relative elongation $\Delta l/l$ along the axis, defined by the guiding cosines $\beta_{x,y,z}$, in a magnetic field the direction of which is determined by the cosines $\alpha_{x,y,z}$, is described by

$$\frac{\Delta l}{l} = \frac{3}{2} \lambda_{100} \left( \alpha_x^2 \beta_x^2 + \alpha_y^2 \beta_y^2 + \alpha_z^2 \beta_z^2 - \frac{1}{3} \right) + 3 \lambda_{111} (\alpha_x \alpha_y \beta_x \beta_y + \alpha_z \alpha_y \beta_z \beta_y + \alpha_x \alpha_z \beta_x \beta_z), \quad (2)$$

where $\lambda_{100}$ and $\lambda_{111}$ are constants of linear magnetostriction in [100] and [111] directions at saturation. In our case, $\Delta l/l$ was measured along the x-axis, so $\alpha_z = \beta_y = \beta_z = 0$, $\beta_x = 1$. Therefore $(\Delta l/l)_{100} = \lambda_{100}$ for $\mathbf{H} \| [100]$ and $(\Delta l/l)_{010} = -\lambda_{100}/2$ for $\mathbf{H} \| [010]$. It is worth to pay attention that for $\alpha_x = 1/\sqrt{3}$, when the angle $\varphi$ between the [100] axis and the direction of magnetization is about 54°, $(\Delta l/l)_{\varphi} = 0$. From Fig. 1b it follows that for $\mathbf{H} \| [100]$ the value of $(\Delta l/l)_{100}$ increases sharply starting at $H = 1.6$ kOe and reaches a saturation value about $-654 \cdot 10^{-6}$ at $H = 2.8$ kOe, i.e., in the same field as the magnetization. Consequently, for our sample, $\lambda_{100}$ = -654·10⁻⁶. The very small value of $(\Delta l/l)_{100}$ for $H$ <1.6 kOe can be explained by an increase in the size of domains with magnetization oriented along the applied field due to corresponding decrease s of domains with opposite magnetization. This process does not change $(\Delta l/l)_{100}$ whereas the total magnetic moment of the sample increases monotonically.

When **H**‖[010], the relative elongation $(\Delta l/l)_{100}$ is positive (Fig. 1c) and reaches about +221·10⁻⁶. This value not two, but three times less than $|(\Delta l/l)_{100}|$ for **H**‖[100], which, like the shape of the $M(H)$ curves, points to some distortion of cubic symmetry.

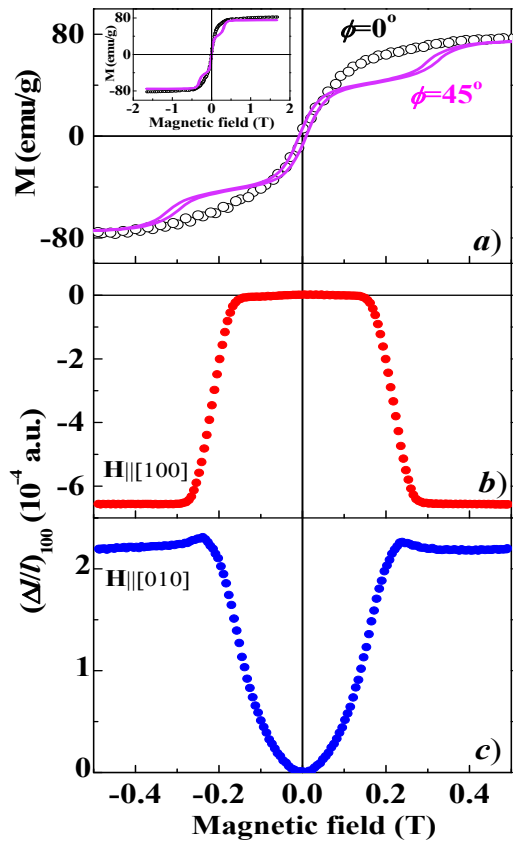

Fig. 1. Magnetic field dependences of *a*) magnetization (*M*) for **H**‖[100] ($\varphi$=0º) and **H**‖[110] ($\varphi$=45º), *b*) and c) magnetostriction $(\Delta l/l)_{100}$ for two magnetic field directions at *T*=295 K. Inset shows full curve of magnetization *M(H)*.

It is generally accepted that the contribution of magnetostriction to the magnetic anisotropy constant $K_1$ does not exceed several percent[13,16,17]. In our case, the situation is different.

One can calculate the contribution of magnetostriction to $K_1$ using the formula[16]

$$\Delta K = \frac{9}{4}[(c_{11} - c_{12})\lambda_{100}^2 - 2c_{44}\lambda_{111}^2]. \quad (3)$$

For the CoFe₂O₄ single crystal, the elastic constants are [18]: $c_{11}$ = 2.57·10¹² erg/cm³, $c_{12}$ = 1.5·10¹² erg/cm³, $c_{44}$ = 0.85·10¹² erg/cm³. Taking $\lambda_{111}$ = 120·10⁻⁶ (from Ref.13) and $\lambda_{100}$ = -654·10⁻⁶, we

obtain $\Delta K \approx 1 \cdot 10^6$ erg/cm$^3$. From the experimental data (Fig.1$a$), the estimated value of $K_1$ is about $2 \cdot 10^6$ erg/cm$^3$. Thus, in the case of CoFe$_2$O$_4$, the contribution of magnetostriction to the magnetic anisotropy constant is abnormally large.

## 4. ABSORPTION SPECTRUM

The spectrum of the light absorption coefficient for the CoFe$_2$O$_4$ single crystal recorded at room temperature is shown in Fig.2a. Close results have been published in Refs.9,19,20. The experimental curve is partially consistent with the optical conductivity spectrum calculated using the Kramers-Kronig analysis of reflectivity spectrum (see Fig.2b and Ref.8). Absorption spectrum $\alpha(\lambda)$ is characterized by a sharp increase at $\lambda < 2$ μm, which is associated with the fundamental absorption edge at $E_g = 1.18$ eV (~1 μm), see Ref. 21. This edge is formed by the indirect transitions from the hybridized $d$Co+$p$O – states of the valence band at $X$ point of the Brillouin zone into the $d$Fe states of the conduction band at $\Gamma$ point [9,22,23]. When the temperature decreases from 400 K to 80 K, the absorption edge undergo a "blue" shift to the region of short wavelengths by +0.08 eV.

In contrast to CoFe$_2$O$_4$, the "red" temperature shift of the absorption edge is observed in the Hg(Cd)Cr$_2$Se$_4$ ferromagnetic spinels. It is due to the strong exchange interaction of charge carriers with localized magnetic moments of Cr ions[24]. Therefore the exchange interaction in CoFe$_2$O$_4$ is substantially lower than that in Hg(Cd)Cr$_2$Se$_4$.

For greater wavelengths, the impurity band ($1$) at $\lambda_1 = 2.6$ μm (0.48 eV) is clearly seen in the spectrum. This band was earlier observed in light transmission[19] at $\lambda = 2.91$ μm and in the reflection spectra[8] at $\lambda = 2.96$ μm. It was assumed in Ref.9 that this band is related to electronic transitions from the valence band into the $V_O$+3d(Fe$^{3+}$) state, where $V_O$ denotes an oxygen vacancy. However, it was recently shown[25] that the oxygen environment of the Co$^{2+}$ and Fe$^{3+}$ ions experiences octahedral distortions, which are stronger in the case of Co$^{2+}$. We can assume that the band ($1$) is formed by the transitions into the $V_O$+3d(Fe$^{3+}$) or/and $V_O$+3d(Co$^{2+}$) states. The decrease in the intensity of band ($1$) with temperature decreasing can be explained by a decrease in the contribution of the "tail" of the absorption edge.

The broad absorption band ($6$) with maximum at $\lambda_6 = 12.5$ μm (0,1 eV) and a fine structure in the wavelength range $3 < \lambda < 15$ μm is seen in Fig. 2$a$. A fine structure is formed by the bands with maxima at $\lambda_2 = 6.1$ μm (0.2 eV), $\lambda_3 = 7$ μm (0.17 eV), $\lambda_4 = 8.4$ μm (0.14 eV), and $\lambda_5 = 10$ μm (0.12 eV). They are close to the impurity absorption bands observed earlier in the optical spectra of polycrystalline CoFe$_2$O$_4$ doped with Zn, Zr or Cd (see, for example, Ref.19). The difference in the spectra of undoped and doped CoFe$_2$O$_4$ is in the enhancement of the absorption background in the region of $3 < \lambda < 15$ μm due to the increase in the concentration of impurities and defectiveness in the cationic sublattice[9,19]. Cooling of the samples leads to "sharpening" of the fine structure of band (*6*).

The further increase in light absorption at $\lambda > 15$ μm is associated with phonons part (Fig. 2*a* and *b*). The phonon spectrum is formed by the $\lambda_{1P}=16.4$ μm ($E_1 = 609$ cm$^{-1}$) band associated with vibrations of the Co – O ions in the octahedral sublattice and the $\lambda_{2P}=24.2$ μm ($E_2 = 413$ cm$^{-1}$) band, associated with vibrations of oxygen in the tetrahedral sublattice [19,20,26,27]. Kramers – Kronig calculations allowed us to identify additional phonon bands at $\lambda_{3P}\approx18.7$ μm ($E = 534$ cm$^{-1}$) and $\lambda_{4P}\approx21.5$ μm ($E = 466$ cm$^{-1}$).

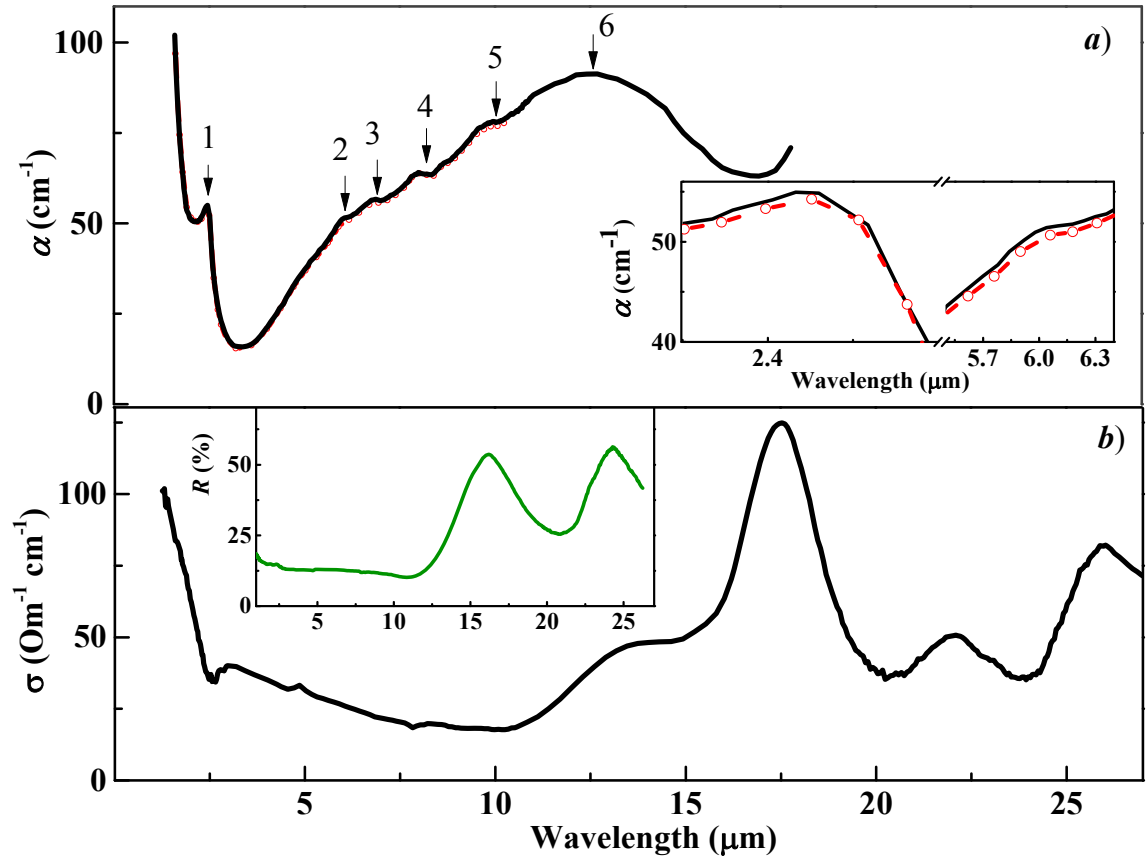

Fig. 2. Spectral dependences of *a)* absorption spectra ($\alpha$*)* of CoFe$_2$O$_4$ at *H*=0 T (solid line) and **H**‖[100]=7.5 kOe (red chain line) at *T*=295 K.The arrows indicate the position of the impurity absorption bands (on inset: enlarged portion of the spectrum) and *b)* optical conductivity ($\sigma$), calculated from the reflection spectrum of CoFe$_2$O$_4$ (see the inset) using Kramers-Kronig analysis.

The magnetic field of *H*=7.5 kOe, which substantially exceeds the saturation value, has an influence upon the absorption $\alpha(\lambda)$. At room temperature, the application of the magnetic field **H**‖[100] leads to a weak "red" shift of the absorption edge, $\Delta E(H) \approx$ -2 meV, in contrast to the "blue shift" $\Delta E(H)\approx$ +10 meV for Faraday experimental geometry[8]. However, when the temperature decreases down to *T* = 80 K in zero magnetic field, the absorption edge is characterized by a "blue shift" [9,22]. Therefore, there is a competition between two mechanisms: temperature-induced "blue shift" vs. magnetic-field-induced "red shift" of the absorption edge.

## 5. EFFECT OF MAGNETIC FIELD ORIENTATION ON THE OPTICAL PROPERTIES

The positions of both the absorption edge and the impurity bands depend not only on the magnitude of the magnetic field applied, but also on its orientation relative to the crystallographic axes. In our case, the magnetization is in the (001) plane. Therefore, the field dependence of the absorption coefficient $\alpha$ for a fully magnetized sample (single domain state) has a simple form:

$$\alpha = A + B sin4\varphi, \qquad (4)$$

where $A$ and $B$ $(|B|<<A)$ are the coefficients that depend on the light wavelength, temperature and magnitude of the magnetic field $H$, $\varphi$ is the angle between the [100] axis and direction of magnetization. If a sample placed in a magnetic field is in a multi-domain state (the magnetic field applied is lower than the saturation field $H_s$), the absorption coefficient can described by the expression $\alpha = A + B\langle sin4\varphi \rangle_H$, where $\langle ... \rangle_H$ means averaging over the domain structure. Then the magnetoabsorption $\Delta\alpha/\alpha$ can be written as

$$\frac{\Delta\alpha}{\alpha} = \frac{\alpha(H) - \alpha(H=0)}{\alpha(H=0)} = \frac{\Delta A + B(\langle sin4\varphi \rangle_H - \langle sin4\varphi \rangle_{H=0})}{A(H=0)}, \qquad (5)$$

where $\Delta A = A(H) - A(H=0)<<A(H=0)$.

Finally, if $\Delta A$ and $B$ are of the same order of magnitude, strong angle dependence of magnetoabsorption, as well as of $\Delta t/t$ and $\Delta R/R$, should be observed.

Fig.3 show the spectra of magneto-transmission $\Delta t/t$, magneto-absorption $\Delta\alpha/\alpha$ and magneto-reflection $\Delta R/R$ for $\mathbf{H}\|[100]$, when the effects reach its maximum values. The spectra have complex shapes with specific features near the absorption bands ($1 - 5$) mentioned before. The shapes and intensities of the features in the $\Delta t/t$ and $\Delta\alpha/\alpha$ spectra are determined by the shift and variations of the intensity of the weak impurity bands upon application of a magnetic field.

Figures 3$a$ and 3$b$ show that the $\Delta t/t$ and $\Delta R/R$ have a positive sign and reach ~ 4 % at $\lambda$>2.2 μm, $\mathbf{H}\|[100]$ and $T$=295 K. As the temperature going down to $T = 80$ K, the magneto-transmission

slightly increase and become positive in the entire measured spectral range and show an additional band at $\lambda$=3.4 μm. This feature was also observed in the magneto-absorption spectrum $\Delta\alpha/\alpha$ (Fig.3c) and in the Faraday rotation of the crystal [9]. Therefore we can contend that the impurity states in the of CoFe₂O₄ crystal are being substantially changed under the influence of either magnetic field or temperature.

For **H**‖[110] variations of $\Delta t/t$, $\Delta\alpha/\alpha$ and $\Delta R/R(H)$ in the spectral region of interest are almost negligible. In accordance with Eq. (5), this fact points to proximity of the values of $A$ and $B$. From Fig.3c we can see that $\Delta\alpha/\alpha$ <0 for $\varphi$=0 (**H**‖[100]), so that we infer $\Delta A$<0 and $B$>0.

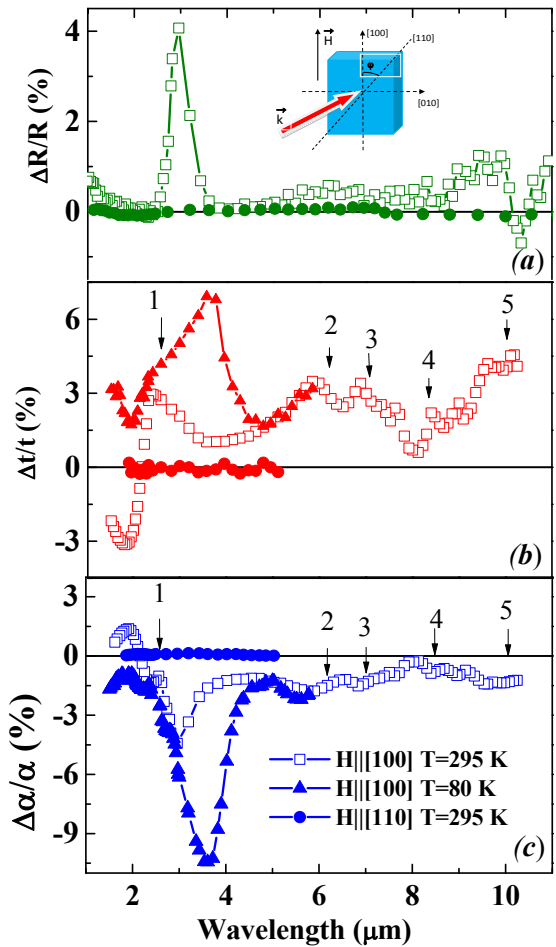

Fig.3. Spectral dependences of *a*) magnetoreflection ($\Delta R/R$), *b*) magnetotransmission ($\Delta t/t$) and *c*) magnetoabsorption ($\Delta\alpha/\alpha$) for the CoFe₂O₄ single crystal at *T*=80 K and 295 K for different orientation of the magnetic field *H*=7.5 kOe (for $\Delta t/t$ and $\Delta\alpha/\alpha$) and 3.6 kOe (for $\Delta R/R$). The arrows indicate the position of the impurity absorption bands. Inset shows the scheme of orientation of the crystal in the in-plane magnetic field.

# 6. MAGNETIC FIELD DEPENDENCES

The most distinct relationship between magnetostriction and magneto-optical properties of the CoFe$_2$O$_4$ crystal can be seen from their magnetic field dependences (Fig.4). The $\Delta R/R(H)$, $\Delta t/t(H)$ and $\Delta\alpha/\alpha(H)$ are the even effects, like magnetostriction, so they are mainly determined by the variations of the diagonal components of the dielectric permittivity tensor. At room temperature the values of $\Delta R/R(H)$, $\Delta t/t(H)$ and $\Delta\alpha/\alpha(H)$ are proportional to the square of magnetization. It was shown[28,29] only a weak contribution of the effect linear in magnetization (namely, the Faraday effect) to the magneto-transmission of the CoFe$_2$O$_4$ crystal can be observed. In the Voight geometry of the experiment, we have found no contribution from the effects linear in magnetization.

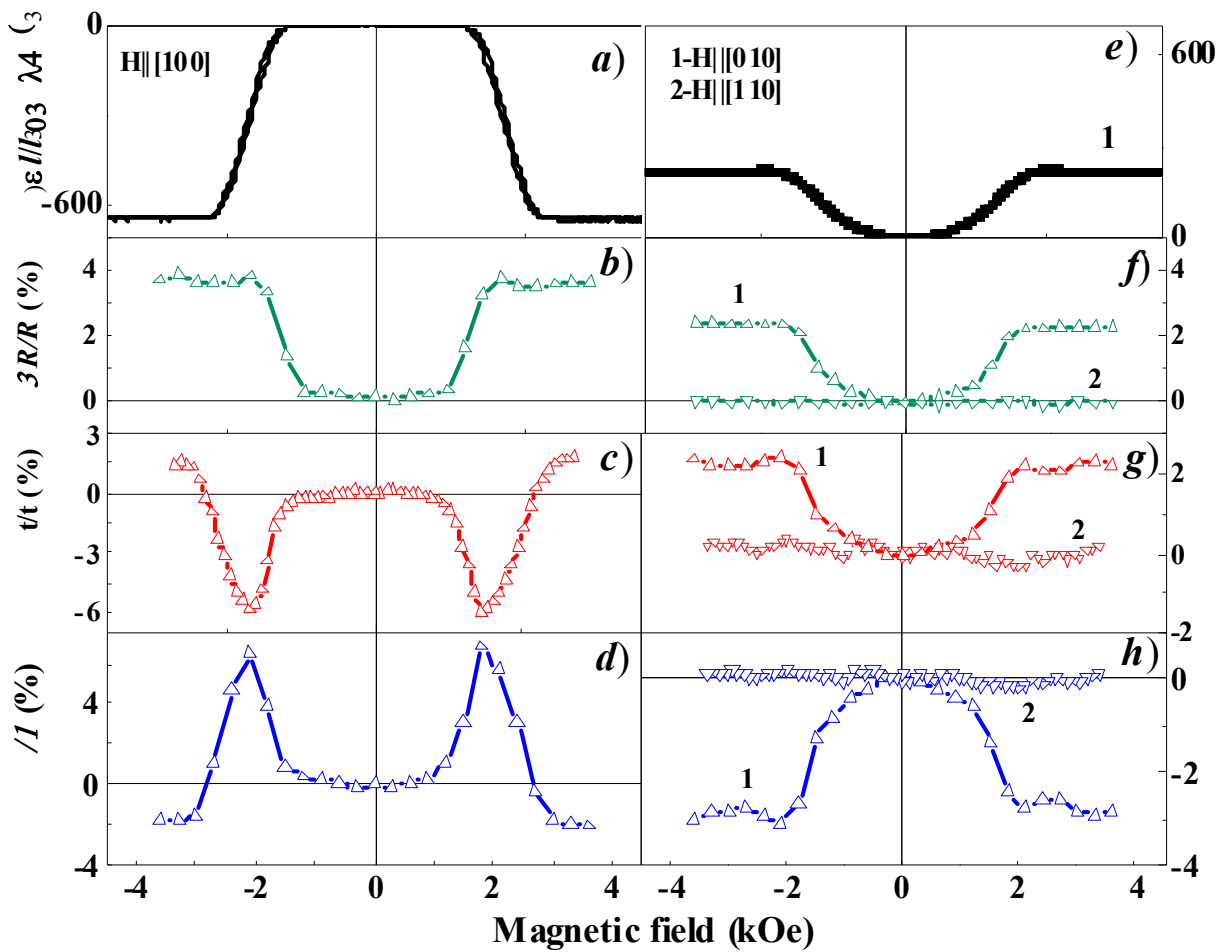

Fig.4. Magnetic field dependences of *a*) linear magnetostriction ($\Delta l/l$)$_{100}$; *b*) magneto-reflection ($\Delta R/R$); *c*) magneto-transmission ($\Delta t/t$) and *d*) magneto-absorption ($\Delta\alpha/\alpha$) of the CoFe$_2$O$_4$ single crystal for orientation of magnetic field **H**||[100] at $\lambda$=2.7 μm and *T*=295. On the right - *e*), *f*), *g*), *h*) are the same dependences, respectively, but in case of **H**||[010] and **H**||[110].

For the Hg(Cd)Cr$_2$Se$_4$ spinels with weak magnetostriction[30], the $\Delta t/t(H)$ and $\Delta R/R(H)$ dependences strictly follow the magnetization[24]. Meanwhile, in the case of the CoFe$_2$O$_4$ spinel with strong magnetostriction, the behavior of the $\Delta t/t(H)$ and $\Delta R/R(H)$ curves are different. Figure 4 demonstrates the $\Delta R/R(H)$, $\Delta t/t(H)$ and $\Delta\alpha/\alpha(H)$ dependences recorded at $\lambda$=2.7 μm, i.e. in the vicinity of the peak (*1*). The shape of curves depends on both the direction of the external magnetic field and spectral position (see, the Supplement for details). For **H**||[100] and H<1.7 kOe, the $\Delta R/R(H)$, $\Delta t/t(H)$ and $\Delta\alpha/\alpha(H)$ are slightly changed in the magnetic field. It is fully consistent with the field dependence of magnetostriction (Fig.4a) and can be explained by the same mechanism. Above 1.7 kOe there is a sharp increase in the magnitude of the magneto-optical effects with saturation at the same fields as for $(\Delta l/l)_{100}$ and magnetization. The extreme on the $\Delta t/t(H)$ and $\Delta\alpha/\alpha(H)$ curves practically coincide with that for the $\frac{d\left(\frac{\Delta l(H)}{l}\right)_{100}}{dH}$ curve, which also indicates the strong connection between magneto-optical effects and magnetostriction in CoFe$_2$O$_4$.

When **H**||[010] there is a smooth growth of $(\Delta l/l)_{100}$ $\Delta R/R(H)$, $\Delta t/t(H)$ and $\Delta\alpha/\alpha(H)$ curves with increasing $H$ but the effects are saturated in the same fields as for the **H**||[100] case. Nevertheless, the values of the magneto-optical effects and $(\Delta l/l)_{100}$ are substantially less: $(\Delta l/l)_{010}$ = $\lambda_{100}/4$ = -1.6·10$^{-4}$ vs. $\lambda_{100}$ = -6.5·10$^{-4}$. The complicated shape of the $\Delta t/t(T,H)$ and $\Delta\alpha/\alpha(T,H)$ curves is probably caused by variations of not only the intensity, but also of the position of the impurity absorption bands.

In the **H**||[110] case, the considered magneto-optical effects are practically absent. Above $H$>2.8 kOe, the sample is fully magnetized, so that $\alpha_x^2 = 1/2$ in formula (2). Assuming $\lambda_{111}$ = +1.2·10$^{-4}$ (from Ref.**Ошибка! Закладка не определена.**) one can obtain from Eq. (2) $(\Delta l/l)_{110}$ = -0.7·10$^{-4}$. It means that for **H**||[110] the deformation of the crystal is almost an order of magnitude smaller than that at **H**||[100]. It could explain such small magnitude of the magneto-optical effects, which is below the accuracy of the experiments.

## 7. CONCLUSION

Investigation of optical and magneto-optical properties of the CoFe$_2$O$_4$ single crystal in the infrared spectral range show the presence of the fine structure of the absorption bands in the "transparency window" associated with impurity states. Surprisingly strong influence of an external magnetic field on the absorption and reflection of natural light is found. The magneto-absorption and magneto-reflection effects substantially depend not only on the magnitude, but also on the orientation of the magnetic field relative to the crystallographic axes. Close relationship between magneto-optical effects and magnetostriction, revealed in the region of impurity absorption, points out to the distortion of the environment of the Co$^{2+}$ and Fe$^{3+}$ ions under an application of a magnetic field as an origin of the observed effects. It is shown experimentally that in the $CoFe_2O_4$ ferrimagnetic spinel magnetostriction gives an abnormally large (about 50%) contribution to the magnetic anisotropy constant $K_1$. In contrast to spinel with low magnetostriction, in the $CoFe_2O_4$ spinel, the influence of the magnetic field on the optical properties is indirect: application of magnetic field results in distortion of the crystal lattice, which, in turn, leads to variations in the electronic structure and optical properties. Therefore, there are different competing mechanisms of magneto-absorption and magneto-reflection effects that can be clarified by comparing its field dependencies at different wavelengths. The discovering of the new mechanism of magneto-absorption of infrared light in magnetostrictive magnetics suppose the further investigations in the area of the so-called "strain-magneto-optics" are being promising for enhancing straintronics` technology.

## ACKNOWLEDGEMENTS

The research was carried out using the equipment of the Collaborative Access Center «Testing Center of Nanotechnology and Advanced Materials» of the IMP UB RAS. The research was carried out within the state assignment of Ministry of Science and Higher Education of the Russian Federation (theme "Spin" No. AAAA-A18-118020290104-2).
Yu.P., A.V., N.G., S.V. and A.P. contributed equally to this work.